\documentclass[aip,
 amsmath,amssymb,
 reprint,%
]{revtex4-1}

\usepackage{graphicx}
\usepackage{dcolumn}
\usepackage{bm}

\usepackage[utf8]{inputenc}
\usepackage[T1]{fontenc}
\usepackage{mathptmx}
\usepackage{etoolbox}
\usepackage[table]{xcolor}
\usepackage[colorlinks=true,linkcolor=blue,citecolor=blue,urlcolor=blue,allbordercolors=white]{hyperref}

\makeatletter
\def\@email#1#2{%
 \endgroup
 \patchcmd{\titleblock@produce}
  {\frontmatter@RRAPformat}
  {\frontmatter@RRAPformat{\produce@RRAP{*#1\href{mailto:#2}{#2}}}\frontmatter@RRAPformat}
  {}{}
}%
\makeatother
\begin{document}


\title{Parametric modeling of mechanical effects on circadian oscillators}
\author{Keith E. Kennedy}
\affiliation{Department of Medicine and Life Sciences, Universitat Pompeu Fabra, Barcelona Biomedical Research Park, 08003 Barcelona, Spain}

\author{Juan F. Abenza}
\author{Leone Rossetti}
\affiliation{Institute for Bioengineering of Catalonia, Barcelona Institute for Science and Technology, 08028 Barcelona, Spain}

\author{Xavier Trepat}
\affiliation{Institute for Bioengineering of Catalonia, Barcelona Institute for Science and Technology, 08028 Barcelona, Spain}
\affiliation{Facultat de Medicina, Universitat de Barcelona, 08036 Barcelona, Spain}
\affiliation{Institució Catalana de Recerca i Estudis Avançats, Barcelona, Spain}

\author{Pablo Villoslada}
\affiliation{Department of Medicine and Life Sciences, Universitat Pompeu Fabra, Barcelona Biomedical Research Park, 08003 Barcelona, Spain}
\affiliation{Hospital del Mar Medical Research Institute, Barcelona Biomedical Research Park, 08003 Barcelona, Spain}

\author{Jordi Garcia-Ojalvo}
\affiliation{Department of Medicine and Life Sciences, Universitat Pompeu Fabra, Barcelona Biomedical Research Park, 08003 Barcelona, Spain}


\begin{abstract}
Circadian rhythms are archetypical examples of nonlinear oscillations.
While these oscillations are usually attributed to circuits of biochemical interactions among clock genes and proteins, recent experimental studies reveal that they are also affected by the cell's mechanical environment.
Here we extend a standard biochemical model of circadian rhythmicity to include mechanical effects in a parametric manner.
Using experimental observations to constrain the model, we suggest specific ways in which the mechanical signal might affect the clock.
Additionally, a bifurcation analysis of the system predicts that these mechanical signals need to be within an optimal range for circadian oscillations to occur.
\end{abstract}

\maketitle

\begin{quotation}
Cells are nonlinear dynamical elements, which in multicellular tissues are commonly coupled to one another.
Much work has been done, both theoretically and experimentally, to understand this coupling and to identify its dynamical consequences from a biochemical viewpoint.
In contrast, much less is known about how the mechanical interactions between cells affect these dynamics.
Recent work has shown, for instance, that circadian oscillations degrade substantially in populations of cells \textit{in vitro} when cell density decreases sufficiently.
Here we use this fact to constrain a standard model of circadian oscillations, and propose a way through which external mechanical signals and internal biochemical interactions could combine in clock cells.
\end{quotation}

\section{Introduction}

Molecular clocks are present in all organisms on Earth, ranging from bacteria to humans \cite{Goldbeter1996}.
One of these clocks is the circadian day/night rhythm, whose intrinsic period is around 24 hours \cite{hastings2003clockwork,Takahashi2017}.
Adhering to this clock is crucial for survival, as organisms need to behave differently in the presence or absence of daylight, in order to thrive in natural ecosystems \cite{kronfeld2003partitioning}.

In multicellular organisms, cells need to synchronize their circadian rhythms across and between tissues \cite{Yamaguchi03,Mohawk2012}.
The general conditions underlying the synchronization of nonlinear systems (of which circadian oscillators are an example) were explored and described in detail by J\"urgen Kurths and others over the 1990s and 2000s \cite{pikovsky2002synchronization,boccaletti2002synchronization,arenas2008synchronization}.
In the case of circadian clocks, rhythm coordination can arise from either biochemical signaling, mechanical interactions, or (most likely) through a combination of both.
While much effort has been devoted to studying the biochemical aspects of circadian cell-cell coordination \cite{ueda2002intercellular,gonze05,bernard2007synchronization,to2007molecular,abraham2010coupling,burckard2022cycle}, the role of mechanical interactions on circadian rhythmicity is still largely unexplored.

Here we address this question by including mechanical factors in a biochemical model based on a standard architecture for genetic oscillations.
The model represents mechanical effects through the activity of the transcription factors YAP/TAZ, which are known to sense mechanical signals \cite{Dupont2011}.
Changes in the cell's microenvironment affect the nuclear localization of these proteins, thereby altering their transcriptional regulatory activity.
Recent experimental observations have revealed that the levels of YAP/TAZ signaling, controlled either through changes in cell density or directly through overexpression, determine the quality of circadian rhythms in mouse fibroblasts \cite{Abenza2022}.
We use this observation, and the measured effects of YAP/TAZ on the clock gene Rev-Erb$\alpha$, to constrain our model and to suggest a potential mechanism through which these mechanical effects may arise.

\section{Experimental constraints}

Circadian oscillations are possible thanks to well structured biochemical networks.
In mammals, the genes at the heart of the clock include the activators Clock and Bmal1, as well as the transcriptional repressors Per and Cry. These genes form a negative feedback loop that is capable of sustaining oscillations in gene expression \cite{Takahashi2017}. The nuclear receptor Rev-Erb$\alpha$ is also known to be involved in the clock mechanism, forming an additional negative feedback loop with Clock and Bmal1 to add robustness to the core network \cite{Cho2012}. 

\begin{figure}[htb]
	\begin{minipage}[t]{0.03\textwidth}
		\vspace{0pt}
		{(a)}
	\end{minipage}
	\begin{minipage}[t]{0.42\textwidth}
		\vspace{0pt}
		\includegraphics[width=\textwidth]{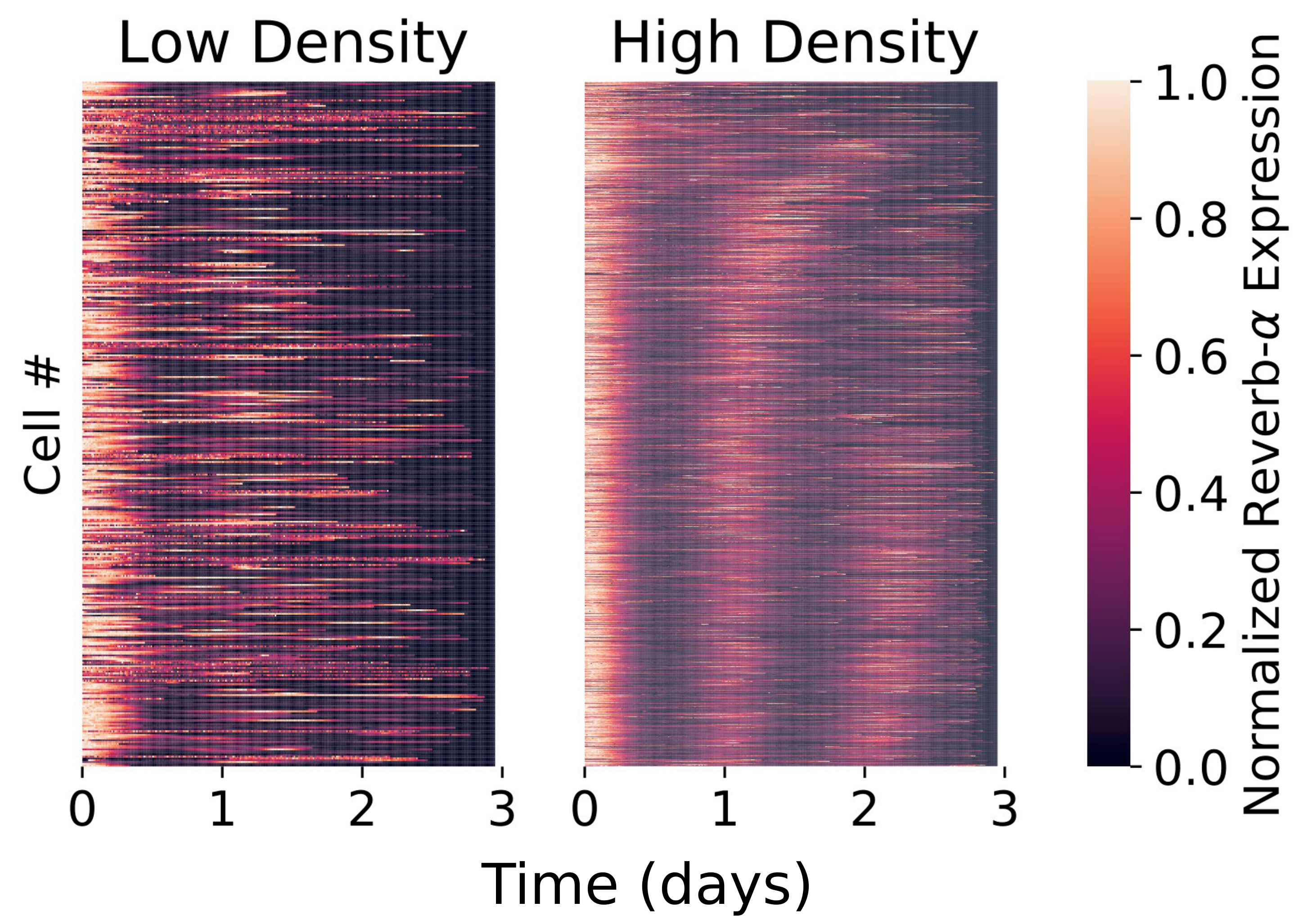}
	\end{minipage}
	\begin{minipage}[t]{0.03\textwidth}
		\vspace{0pt}
		{(b)}
	\end{minipage}
	\begin{minipage}[t]{0.42\textwidth}
		\vspace{0pt}
		\includegraphics[width=\textwidth]{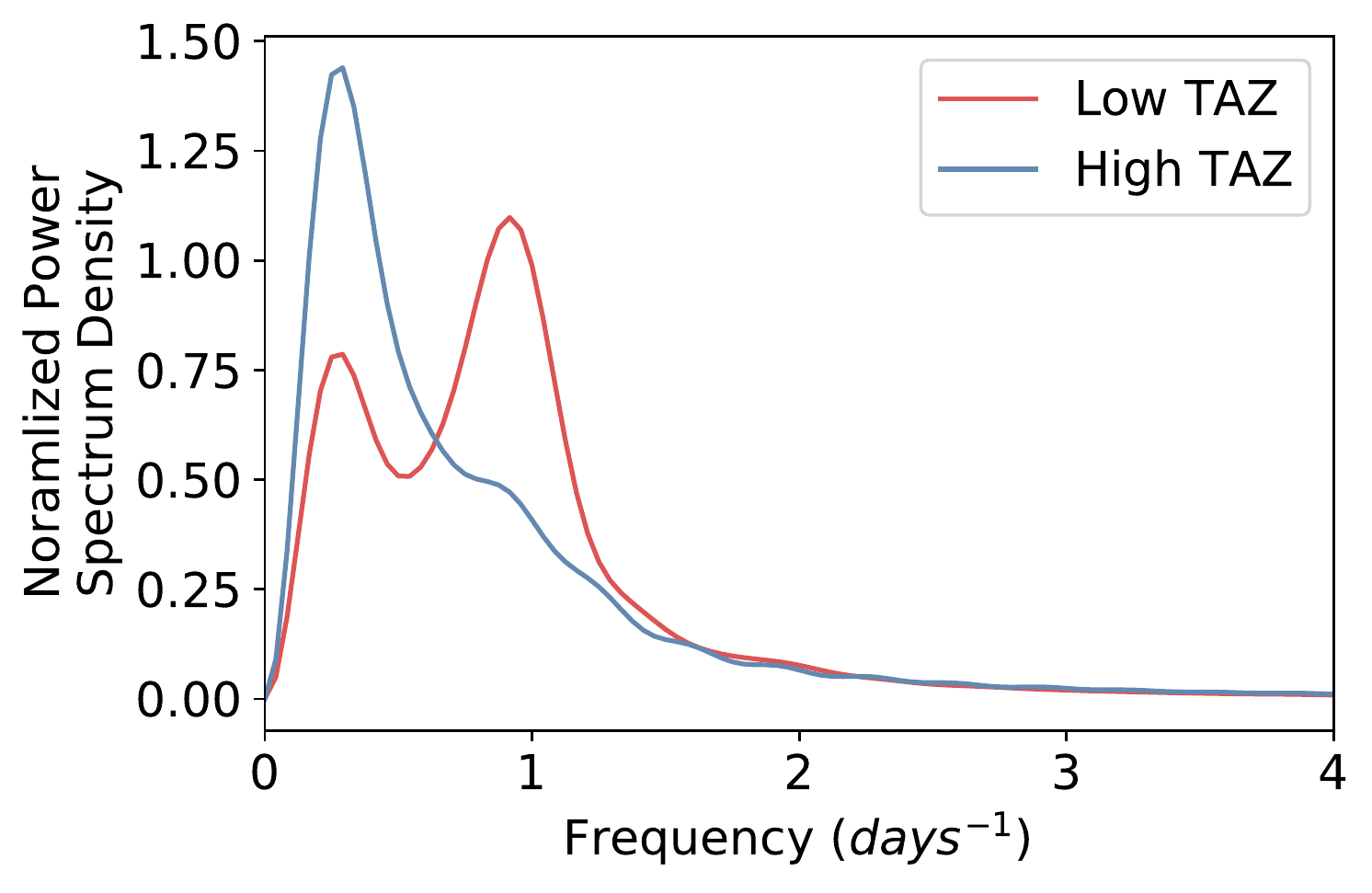}
	\end{minipage}
	\begin{minipage}[t]{0.03\textwidth}
		\vspace{0pt}
		{(c)}
	\end{minipage}
	\begin{minipage}[t]{0.42\textwidth}
		\vspace{0pt}
	\includegraphics[width=0.9\textwidth]{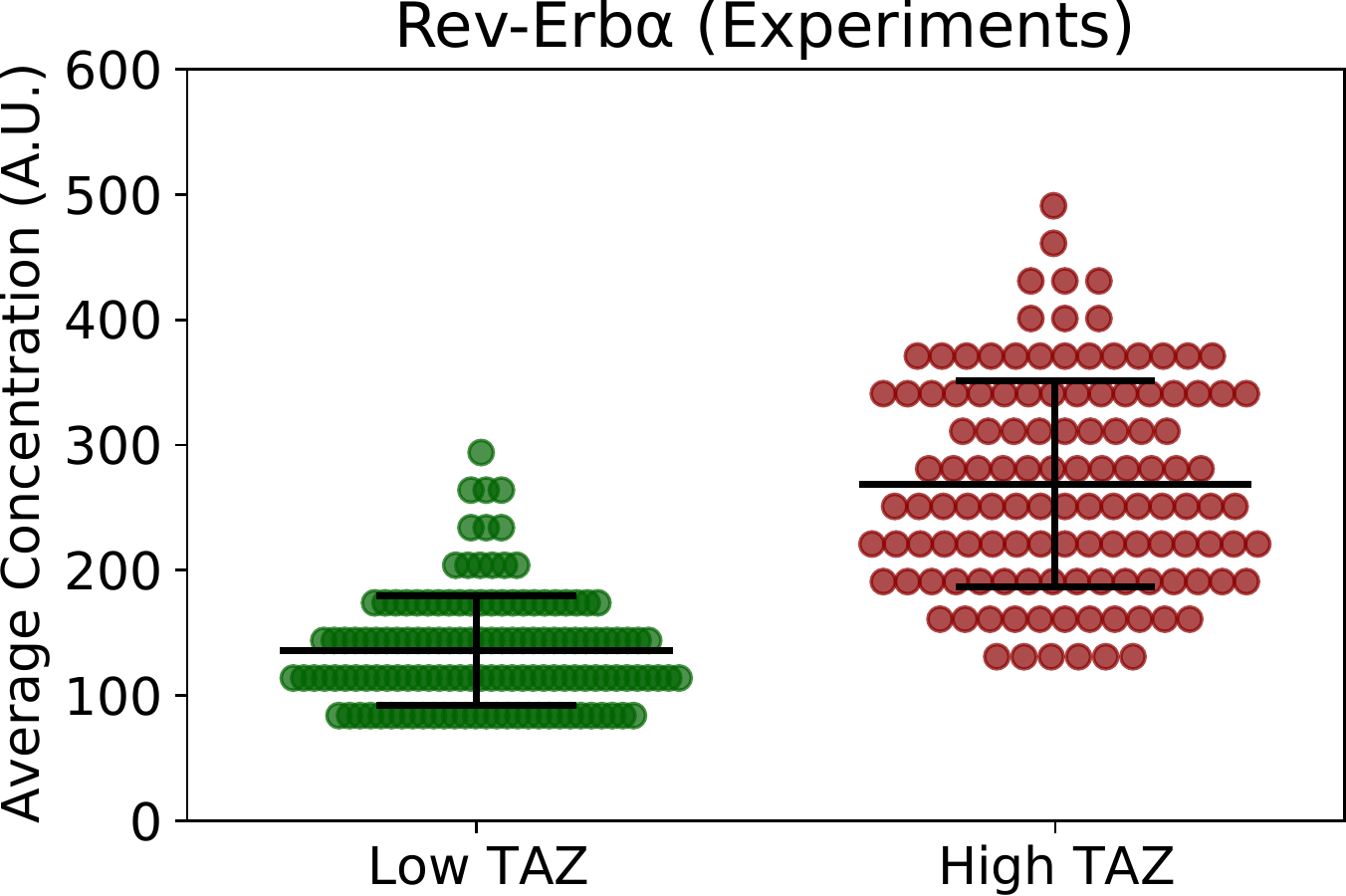}
	\end{minipage}
\caption{Experimental results used to constrain our model.
(a) Kymographs of Rev-Erb$\alpha$ expression in low (left) and high (right) densities mouse fibroblasts in \textit{in vitro}.
Each experiment lasted 72 hours, and a total of 300 time points were sampled.
The horizontal traces correspond to single cells, and are ordered vertically by increasing amplitude of their circadian frequency.
(b) Averaged Fourier spectra for two different levels of TAZ expression.
(c) Effect of TAZ levels on Rev-Erb${\alpha}$ expression.
Data adapted from Abenza et al \cite{Abenza2022}.}
\label{fig:exp}
\end{figure}

Recent experiments in mouse fibroblasts \cite{Abenza2022} have shown that Rev-Erb$\alpha$ is cell-density dependent, which results in circadian rhythms being more robust for higher cell densities.
A representative realization of those experimental results is shown in Fig.~\ref{fig:exp}(a).
This behavior has physiological consequences: during wound healing, for instance, circadian robustness is lost as the cells on the wound front have lower cell density \cite{Abenza2022}.
The molecular connection between cell-cell contact and the molecular clock is believed to be mediated by YAP and TAZ, whose concentrations are known to decrease with increasing cell density \cite{Hsiao16}. In agreement with this fact, overexpression of TAZ leads to a disappearance of circadian rhythmicity, as shown in Fig.~\ref{fig:exp}b (blue line). This contrasts with the behavior exhibited by the cells in basal conditions, for which a clear spectral peak at a period of 1 day is observed (red line in Fig.~\ref{fig:exp}b).
That circadian peak is completely absent for the case of TAZ overexpression, which corresponds to low cell densities in the experiments shown in Fig.~\ref{fig:exp}(a).
Additionally, experiments show that TAZ overexpression leads to increased Rev-Erb$\alpha$ levels (Fig.~\ref{fig:exp}c).

\section{Modeling the circadian effects of YAP/TAZ}
\label{sec:model}

The molecular mechanism through which increased YAP/TAZ nuclear concentration (and correspondingly low cell density) causes cells to lose their circadian rhythmicity remains unknown.
In what follows, we address this question using a variation of the Goodwin model to simulate the effect of YAP/TAZ on the robustness of the circadian clock.
Our model aims to propose a specific mechanism by which YAP/TAZ can affect the quality of circadian oscillations.
Furthermore, the model aims to predict possible outcomes that were not tested in previous experiments, such as the effect of decreasing YAP/TAZ levels (and correspondingly increasing cell density).
Also, it will help us to better understand how the system responds to continuous changes in YAP/TAZ levels: Do cells lose the oscillations suddenly at a certain concentration of YAP/TAZ, or do the oscillations gradually fade away?

The model aims to simplify as much as possible the interactions among the molecular species present in the system, keeping only the essential elements of the clock.
We constrain the model using the experimental observations described above, together with scaling information such as the period of oscillation and typical molecule levels of the clock components.
 
The Goodwin model \cite{Goodwin1965} is commonly used to describe the dynamics of circadian oscillators \cite{ullner09,Ananthasubramaniam2020}.
It consists of three coupled ordinary differential equations describing the dynamics of three biochemical components connected to one another in a negative feedback loop:
\begin{eqnarray}
	&\frac{dX}{dt}& = \frac{\alpha_1}{1+(\frac{Z}{K}) ^{h}} - d_1X, \\
	&\frac{dY}{dt}& = \beta_2X - d_2Y, \\
	&\frac{dZ}{dt}& = \beta_3Y - \frac{d_3Z}{1+\frac{Z}{S}}.
	\label{eqn:goodwin}
\end{eqnarray}
Here $\alpha_1$ and $K$ are the overall strength and threshold, respectively, of the negative feedback of $Z$ on $X$, $h$ is the Hill coefficient, $\beta_1$ and $\beta_2$ are production rates, and $d_1$, $d_2$, and $d_3$ are decay rates.
This model can be used to simulate the expression levels of a simplified version of the molecular clock network, in which only the feedback loop among Bmal1 mRNA (represented by $X$ above), Rev-Erb$\alpha$ mRNA ($Y$), and Rev-Erb$\alpha$ protein ($Z$) is considered \cite{Ananthasubramaniam2020}. 

An advantage of the Goodwin model is its simplicity and well-documented use. However, the equations must be  adjusted to realistically represent a circadian clock.
First, saturation was added to the decay term of Rev-Erb$\alpha$ protein ($Z$) in Eq.~\eqref{eqn:goodwin} to increase the modulation depth of the oscillations of that variable, which was too small when using the standard parameter set in the common version of the model with linear saturation \cite{Ananthasubramaniam2020}.
Additionally, the expression parameters must be properly selected to reflect known concentrations of molecules within a cell.
Furthermore, the time scale must also be adjusted to produce a period on the order of 24 hours.
This can be done by rescaling the model variables as usually done in dimensional analysis, and selecting the scaling factors that lead to the correct ranges in the values of the variables.
These factors also rescale the model parameters, leading to parameter values that produce oscillations with biologically realistic period and amplitude ranges \cite{Milo2008}.
The behavior of the rescaled model can be seen in Fig.~\ref{fig:goodwin_scaled} below, and the corresponding parameter values are given in Table~\ref{tab:param_vals}.

\begin{figure}[htbp]
	\centering
	\hfill
	\includegraphics[width=0.15\textwidth]{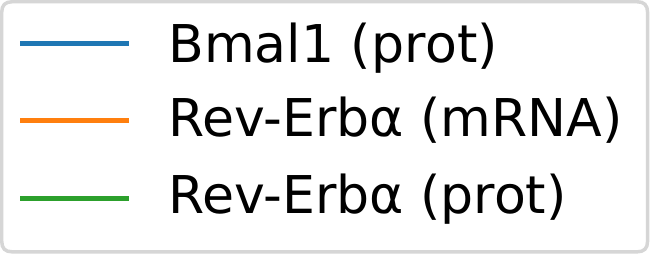}\\
	\includegraphics[width=0.45\textwidth]{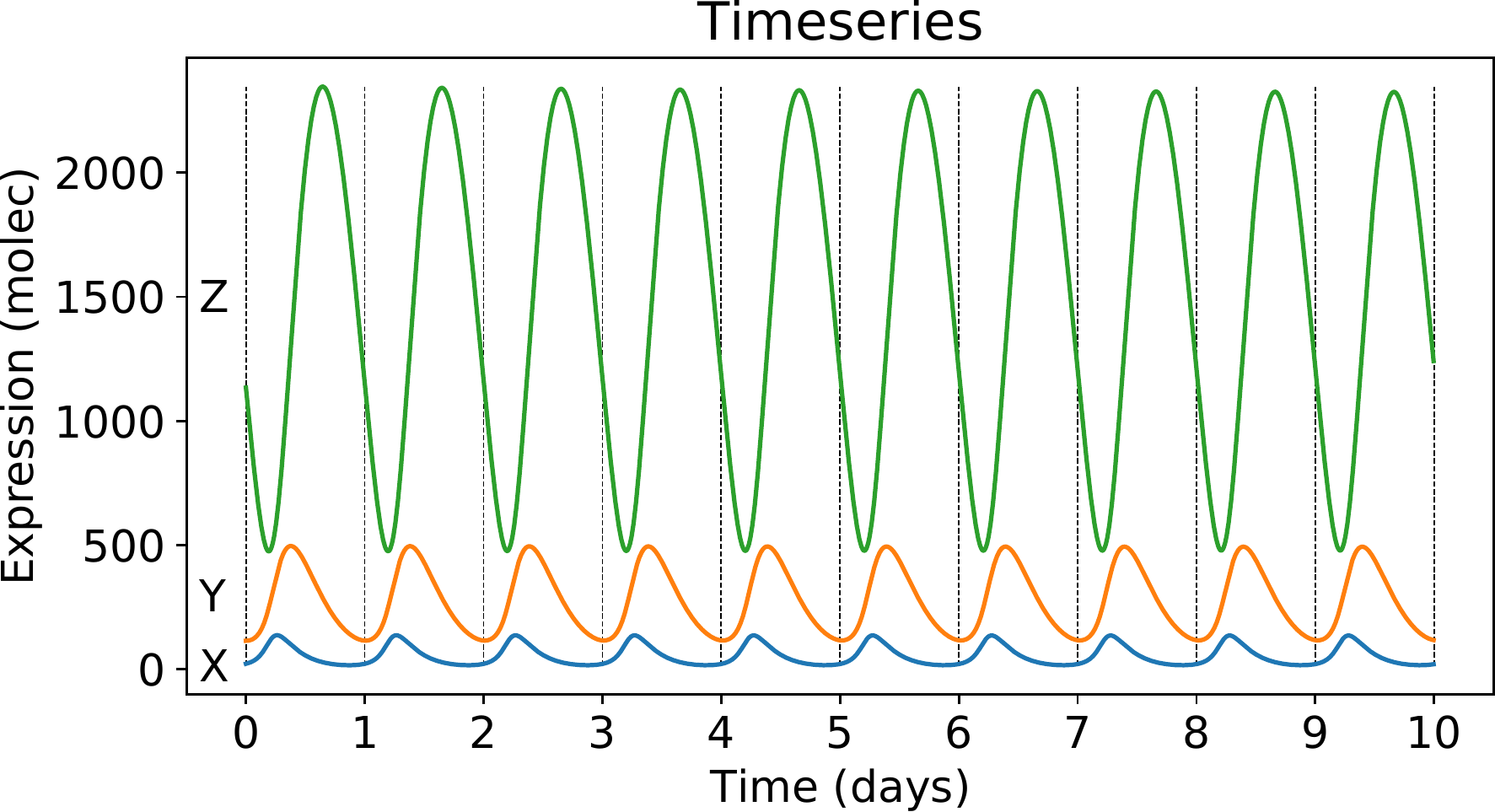}
	\caption{Sample time series for the deterministic model with added saturation.
	Parameters are those given in Table~\ref{tab:param_vals}.}
	\label{fig:goodwin_scaled}
\end{figure}

\begin{table}[htbp]
\centering
\begin{tabular}{l | c | c | c}
	Function                               & Parameter        & Value 	& Units \\ \hline
	Negative Feedback Strength 	  & $\alpha_1$       & 5500    & molec$\cdot$h$^{-1}$ \\  \hline

	Production                           & $\beta_2$         & 1.3       & h$^{-1}$ \\ 
	                                            & $\beta_3$         & 1.3       & h$^{-1}$ \\  \hline

	Degradation                         & $d_1$              & 0.26      & h$^{-1}$ \\ 
					    & $d_2$             & 0.26      & h$^{-1}$ \\ 
                                              & $d_3$              & 3.9       & h$^{-1}$ \\  \hline

   Hill Coefficient                       & $h$              & 2          & - \\  \hline

   Negative Feedback Threshold & $K$              & 50        & molec \\  \hline

	Saturation                            & $S$              & 100       & molec \\  \hline
	\end{tabular}
	\caption{Parameter values used for the deterministic and stochastic simulations.}
	\label{tab:param_vals}
\end{table}

As mentioned above, experiments show that Rev-Erb$\alpha$ levels increase in concentration with TAZ expression (Fig.~\ref{fig:exp}c) \cite{Abenza2022}, both in its mean and standard deviation.
The model therefore needs a way to introduce YAP/TAZ (and thus cell density) that leads to a comparable effect.
On the basis of previous experimental evidence suggesting that YAP/TAZ affects the clock protein Bmal1 \cite{Zhao2008,Zanconato2015,Lee2016,Rivera-Reyes2018,Rajbhandari2018}, and after considering several alternatives, we found that the best agreement with the experimental behavior shown in Fig.~\ref{fig:exp} was obtained when the effect of YAP/TAZ in the clock was represented parametrically through modulation of the negative feedback strength $\alpha_1$.
Specifically, we consider in what follows that an increase in $\alpha_1$ corresponds to a larger inhibitory activity of YAP/TAZ on Bmal1 (and correspondingly to a lower cell density).
As we show in what follows, increasing $\alpha_1$ disrupts the oscillations seen in the system, and increases the average concentration of Rev-Erb$\alpha$ \cite{Abenza2022}.

\begin{figure}[htbp]
	\centering
	\includegraphics[width=0.47\textwidth]{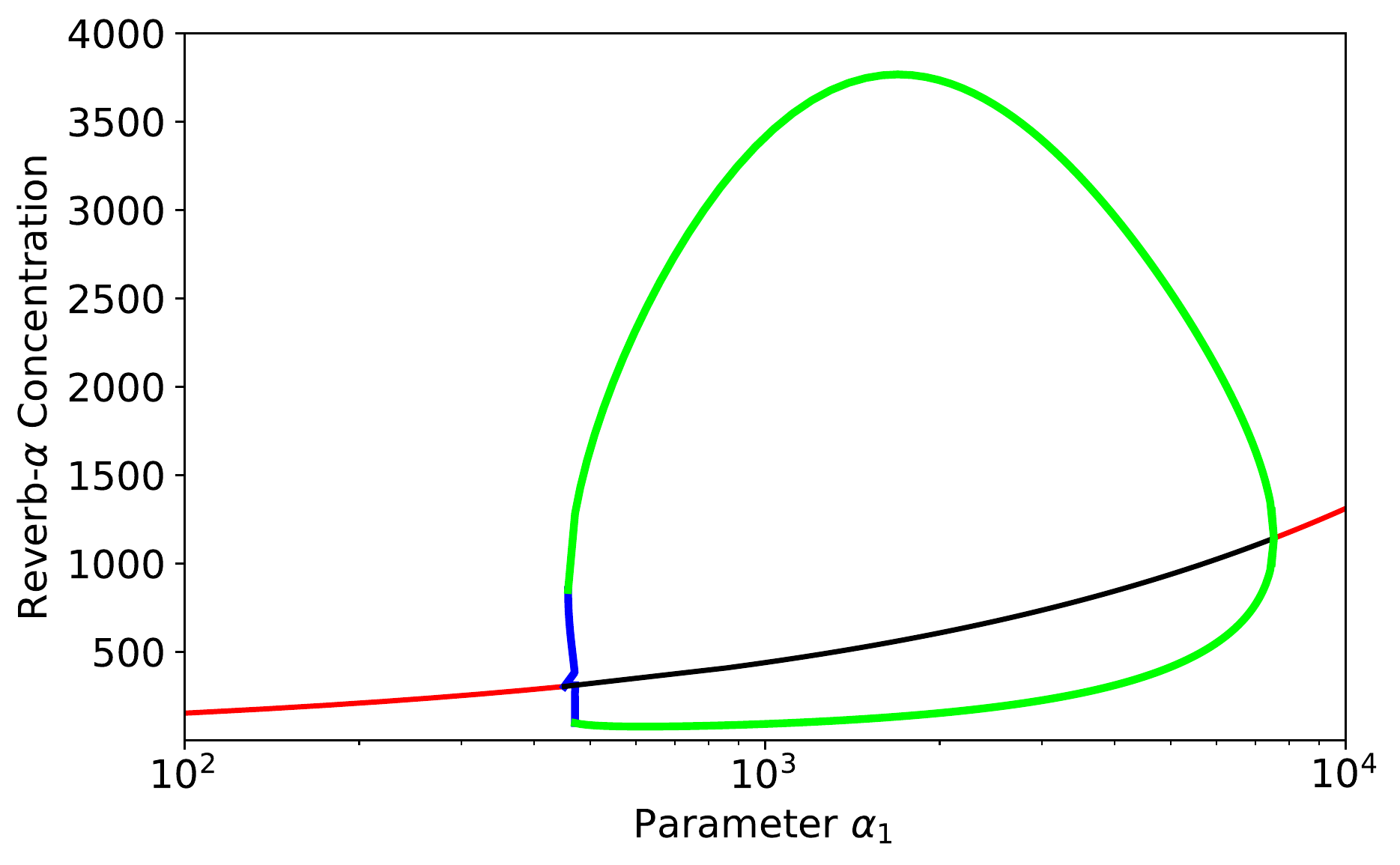}
	\includegraphics[width=0.47\textwidth]{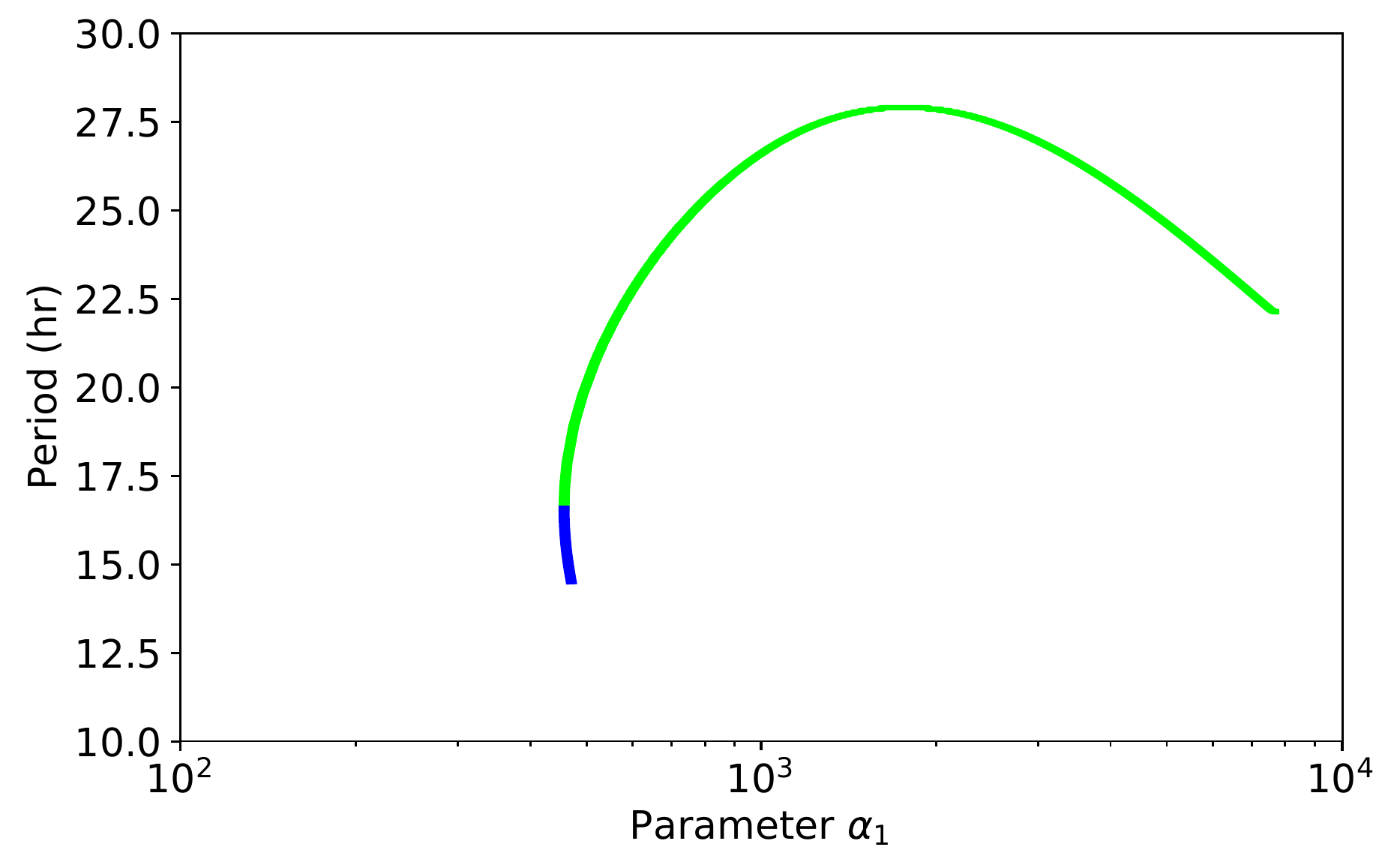}
	\caption{Bifurcation diagram of the extended Goodwin model with saturation.
	Effect of changing the negative feedback parameter $\alpha_1$ on the concentration of Rev-Erb$\alpha$ (top) and the period (bottom).
	Red (black) lines indicate a stable (unstable) equilibrium, and green (blue) lines denote a stable (unstable) limit cycle.
	The maximum period occurs at the same point as the maximum amplitude. The system cycles with a period of 24 hours for $\alpha_1=5500$~molec$\cdot$h$^{-1}$, and this value is used as the base for the model below.}
	\label{fig:bifurcation_model}
\end{figure}

To establish how the nature of the oscillations changes with $\alpha_1$ we use bifurcation analysis.
The results of this analysis, performed with XPPAUT \cite{xppaut_book} and shown in Fig.~\ref{fig:bifurcation_model}, reveal that the system undergoes two Hopf bifurcations, one subcritical at $\alpha_1 \sim 470$~molec$\cdot$h$^{-1}$ and the other supercritical at $\alpha_1 \sim 7500$~molec$\cdot$h$^{-1}$.
At the subcritical point, the system changes from settling to a stable steady state to undergo oscillations. At the supercritical point the reverse occurs: the system stops oscillating and returns to a stable steady state.
The amplitude varies within the region of oscillations, reaching a maximum of around $1800$~molec at $\alpha_1 \sim 1700$~molec$\cdot$h$^{-1}$, and steadily decreasing until reaching the upper Hopf point. 
The period varies somewhat, with a minimum of $\sim 17$~h at the lower Hopf point, rapidly increasing to a maximum value of $\sim 27$~h at $\alpha_1\sim 1700$~molec$\cdot$h$^{-1}$, then steadily decreasing to $\sim 22$~h at the upper Hopf point.

It is also worth pointing out that according to the bifurcation analysis results shown in Fig.~\ref{fig:bifurcation_model}, the fixed point of the system increases monotonically with $\alpha_1$ (both when stable and when unstable).
This fact, together with the behavior of the extrema of the limit cycle itself, allow us to infer that the average Rev-Erb$\alpha$ levels increase monotonically with YAP/TAZ levels, in agreement with the experimental observation (Fig.~\ref{fig:exp}c).

\section{Stochastic modeling}

Deterministic simulations were run first, as described above, to determine the proper values for parameters and establish a reasonable scale for molecule numbers.
They provide a good starting point for the model, and allow us to make qualitative inferences about the system.
However, they lack the noise seen in true biological systems, since the same time series is always produced from a given initial condition.
Stochastic simulations more accurately capture biological processes through the addition of randomness. 
The next reaction method, a type of stochastic simulation, was used to simulate the circadian system, as implemented in the Python package \texttt{StochPy} \cite{Maarleveld2013}.
The next reaction method considers each process in the system as a separate biochemical reaction that has a certain likelihood to occur, determined by its respective propensity \cite{gillespie07}.
In our case, the reactions are either the creation or degradation of one of the components of system, and the propensities for each are calculated using the same parameters from the deterministic model.

\begin{figure*}[htbp]
	\centering
	\begin{minipage}[t]{0.8\textwidth}
	{\large (a) High Density ($\alpha_1 = 5500$~molec$\cdot$h$^{-1}$)}
\hfill
	\includegraphics[width=0.3\textwidth]{figs/fig4a.pdf} \\
	\end{minipage}
	\vskip3mm
	\begin{minipage}[c]{0.8\textwidth}
	\includegraphics[width=\textwidth]{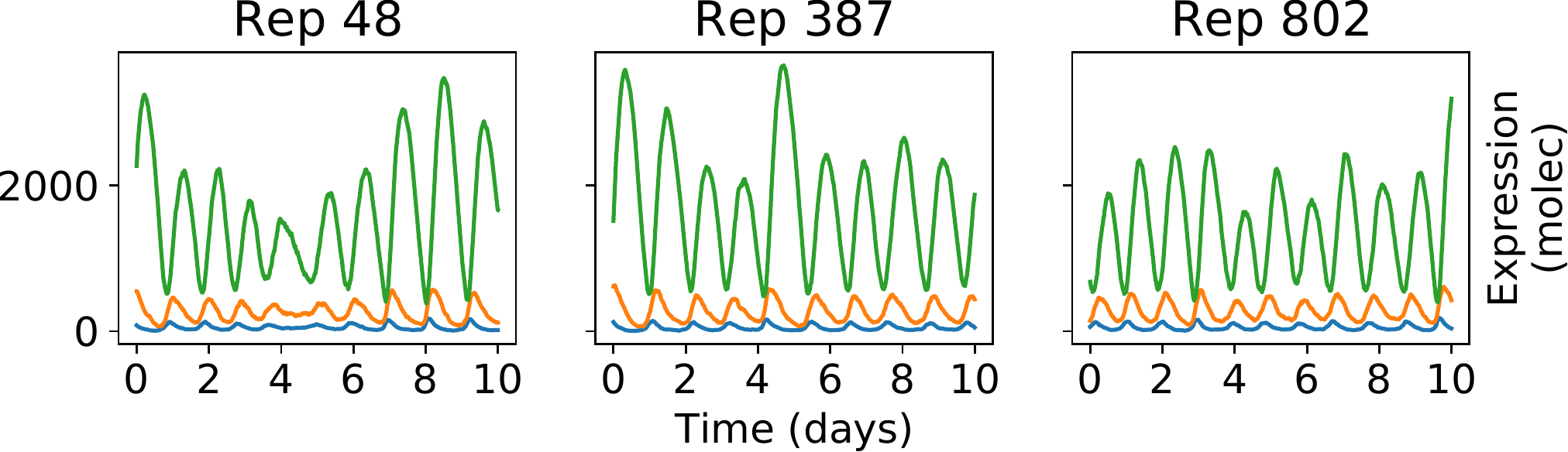}
	\includegraphics[width=\textwidth]{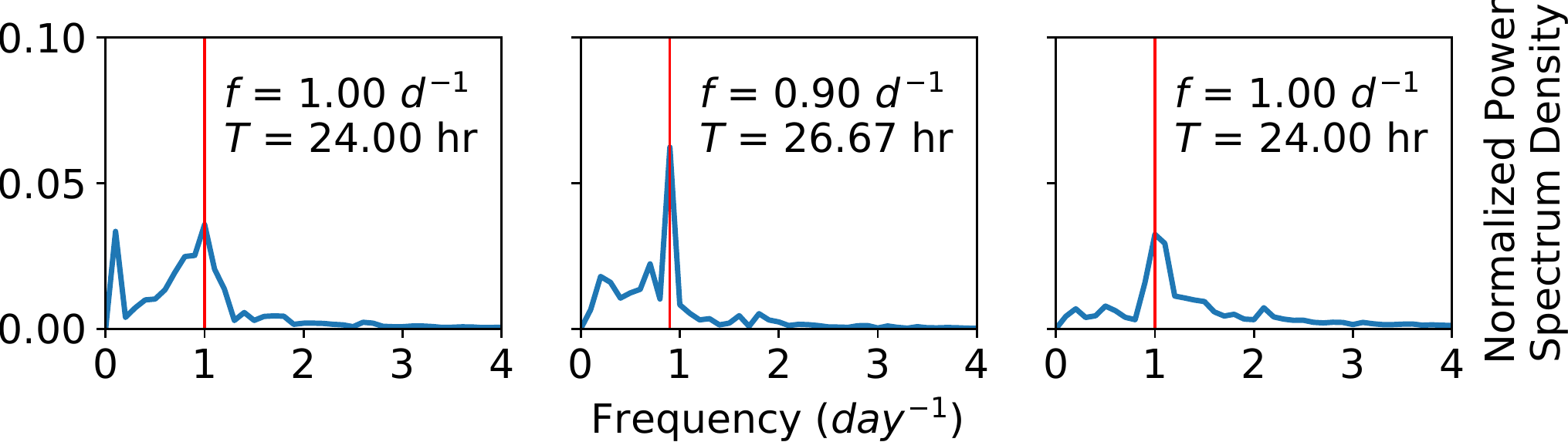} \\
	\end{minipage}
	\vskip3mm
	\begin{minipage}[c]{0.8\textwidth}
	{\large (b) Low Density ($\alpha_1 = 10^5$~molec$\cdot$h$^{-1}$)} \hfill \mbox{}
	\end{minipage}
	\vskip3mm
	\begin{minipage}[c]{0.8\textwidth}
	\includegraphics[width=\textwidth]{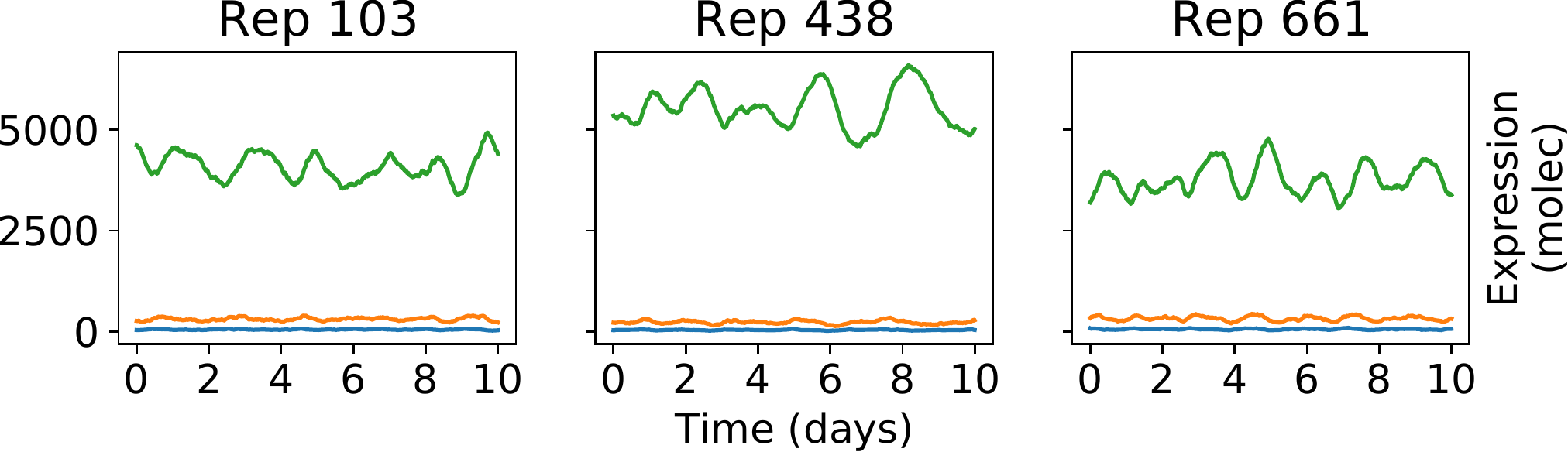}
	\includegraphics[width=\textwidth]{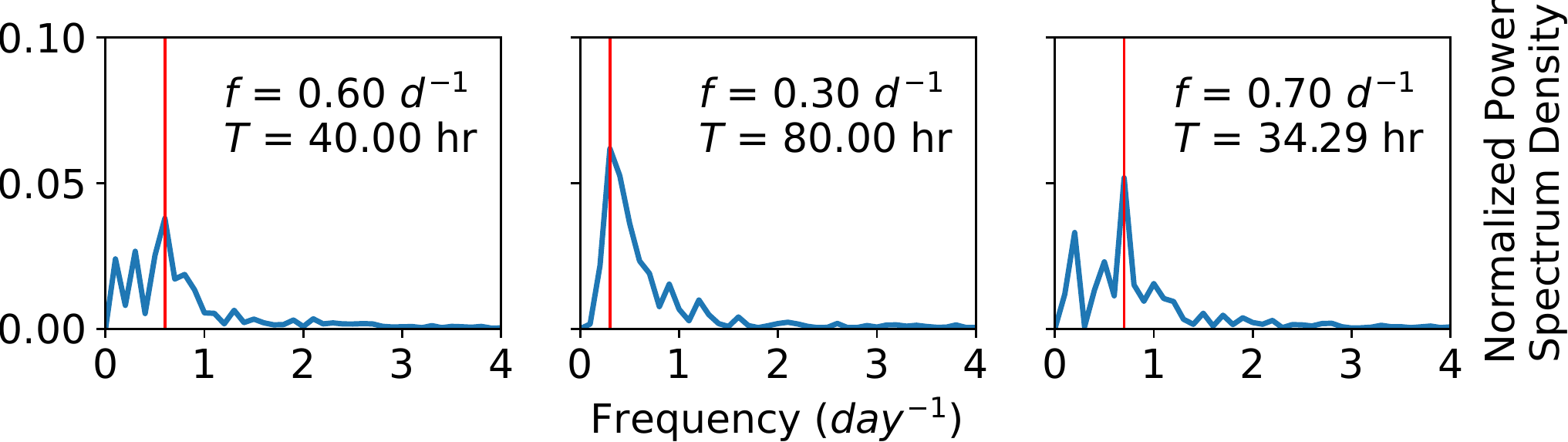}
	\end{minipage}
	\begin{minipage}[c]{0.8\textwidth}
	\caption{Sample time series and Fourier spectra from the stochastic simulations. Three samples from the 1000 repetitions are shown for each of the high and low density cases.}
	\label{fig:sample_timeseries}
	\end{minipage}
\end{figure*}

\begin{figure}[htbp]
	\begin{minipage}[t]{0.03\textwidth}
		\vspace{0pt}
		{(a)}
	\end{minipage}
	\begin{minipage}[t]{0.42\textwidth}
		\vspace{0pt}
		\includegraphics[width=\textwidth]{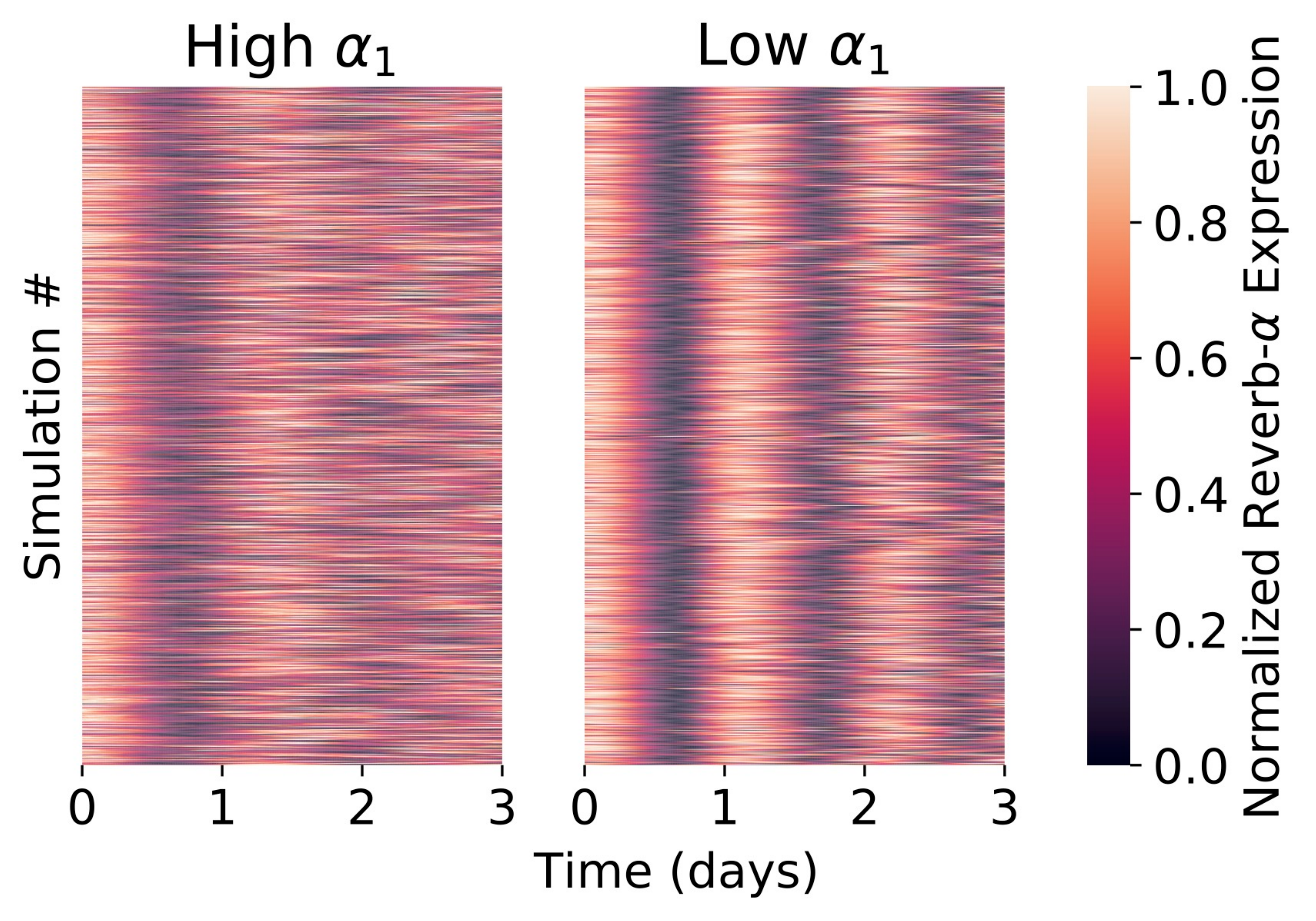}
	\end{minipage}
	\begin{minipage}[t]{0.03\textwidth}
		\vspace{0pt}
		{(b)}
	\end{minipage}
	\begin{minipage}[t]{0.42\textwidth}
		\vspace{0pt}
		\includegraphics[width=\textwidth]{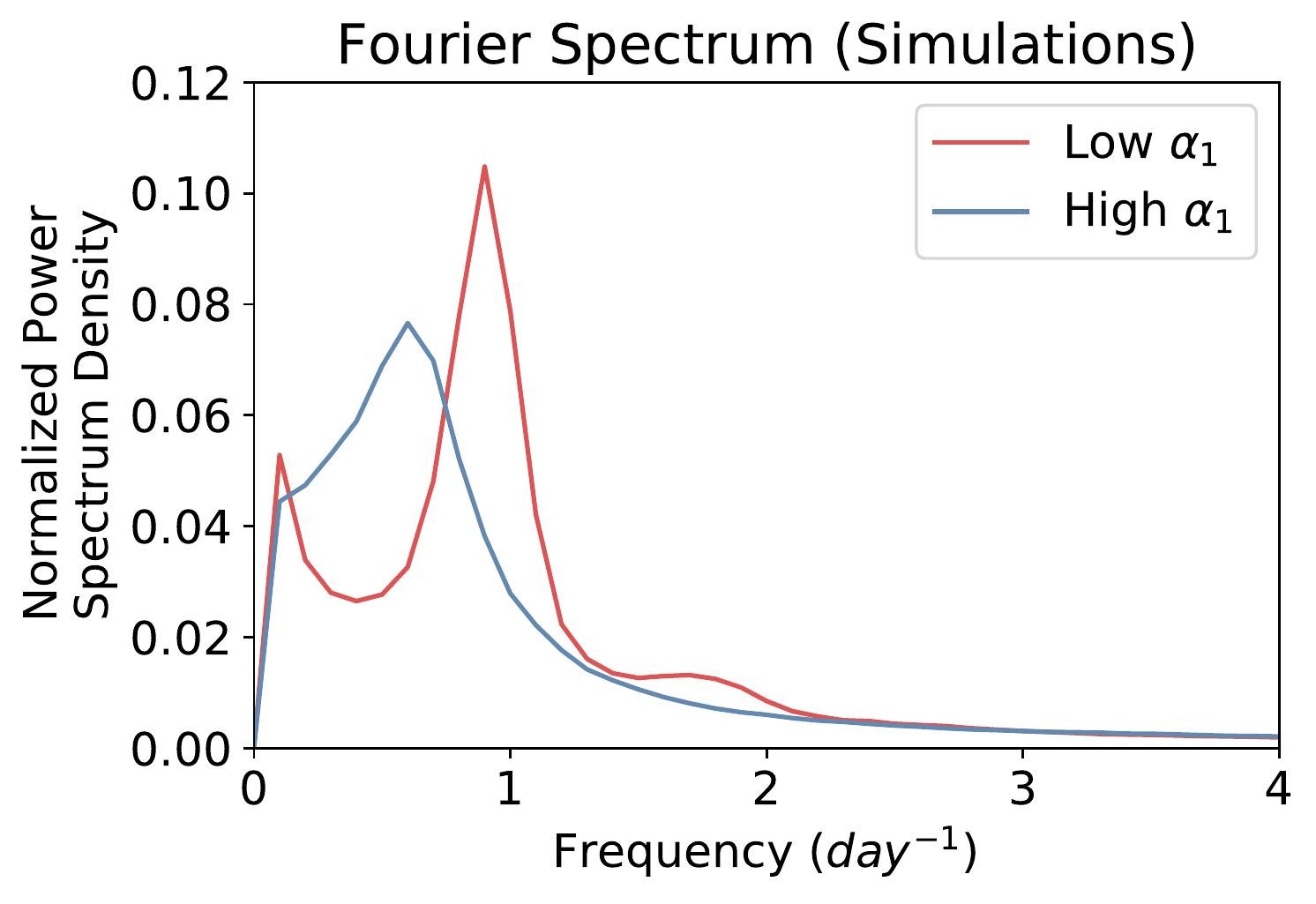}
	\end{minipage}
	\begin{minipage}[t]{0.03\textwidth}
		\vspace{0pt}
		{(c)}
	\end{minipage}
	\begin{minipage}[t]{0.42\textwidth}
		\vspace{0pt}
	\includegraphics[width=0.9\textwidth]{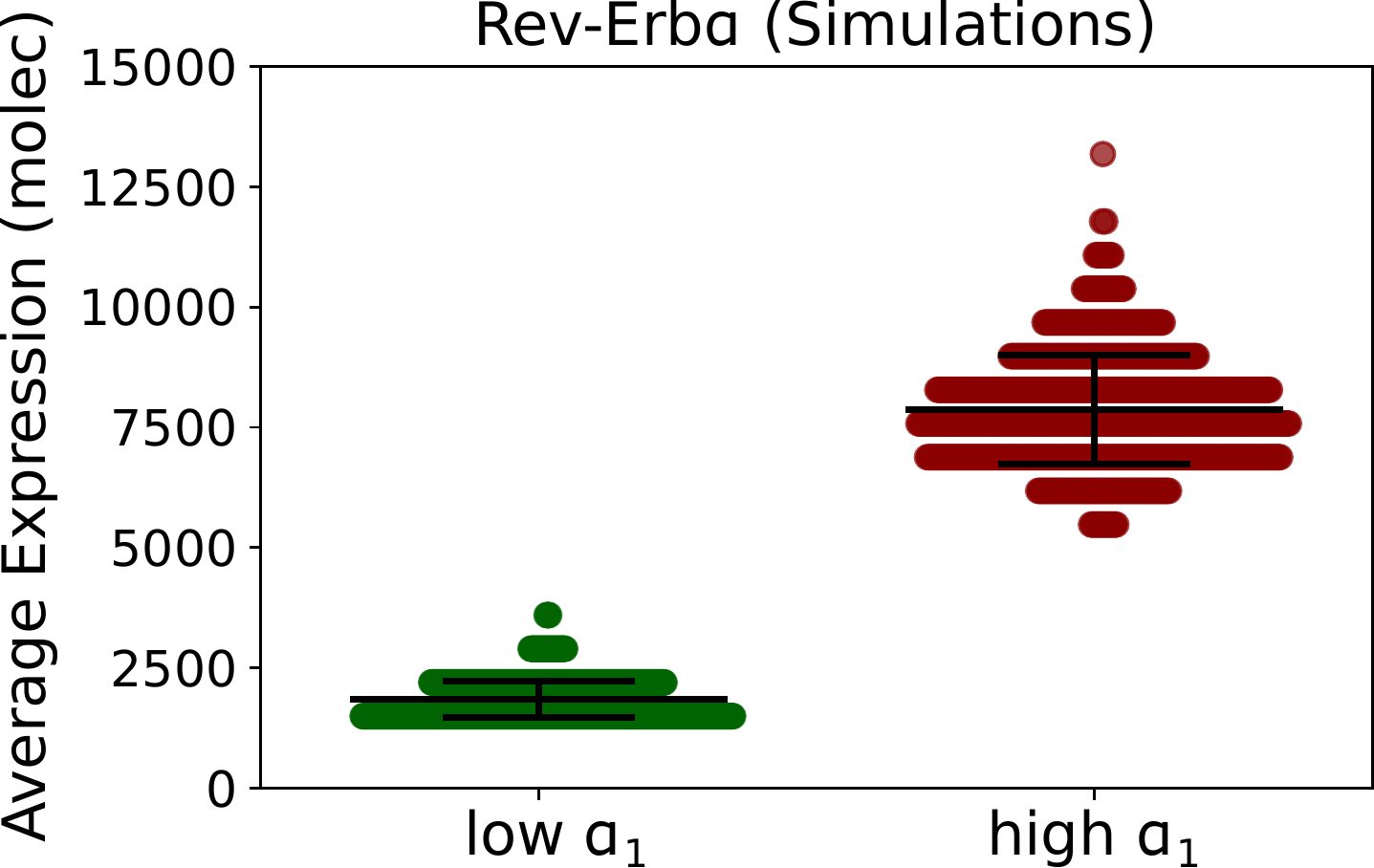}
	\end{minipage}
	\caption{(a) Kymograph of time series obtained with the stochastic simulations. Results are shown for both high ($\alpha_1 = 5500$~molec$\cdot$h$^{-1}$) and low ($\alpha_1 = 10^5$~molec$\cdot$h$^{-1}$) values of the feedback strength, corresponding to low and high cell densities as shown in the experiments of Fig.~\ref{fig:exp}(a).
 As in that figure, the horizontal traces (1000 in each case) are ordered vertically by increasing amplitude of their circadian frequency.
	The time series are shown after a transient period, and are aligned by their first peak in expression.
	(b) Fourier spectrum averaged over 1000 stochastic simulations for the two values of $\alpha_1$ given above.
	The averages were taken over 1000 stochastic simulations each.
	(c) Effect of YAP/TAZ concentration on average concentration of Rev-Erb${\alpha}$ in the stochastic simulations.}
	\label{fig:kymograph_simulations}
\end{figure}

\begin{figure}[htbp]
	\centering
	\includegraphics[width=0.47\textwidth]{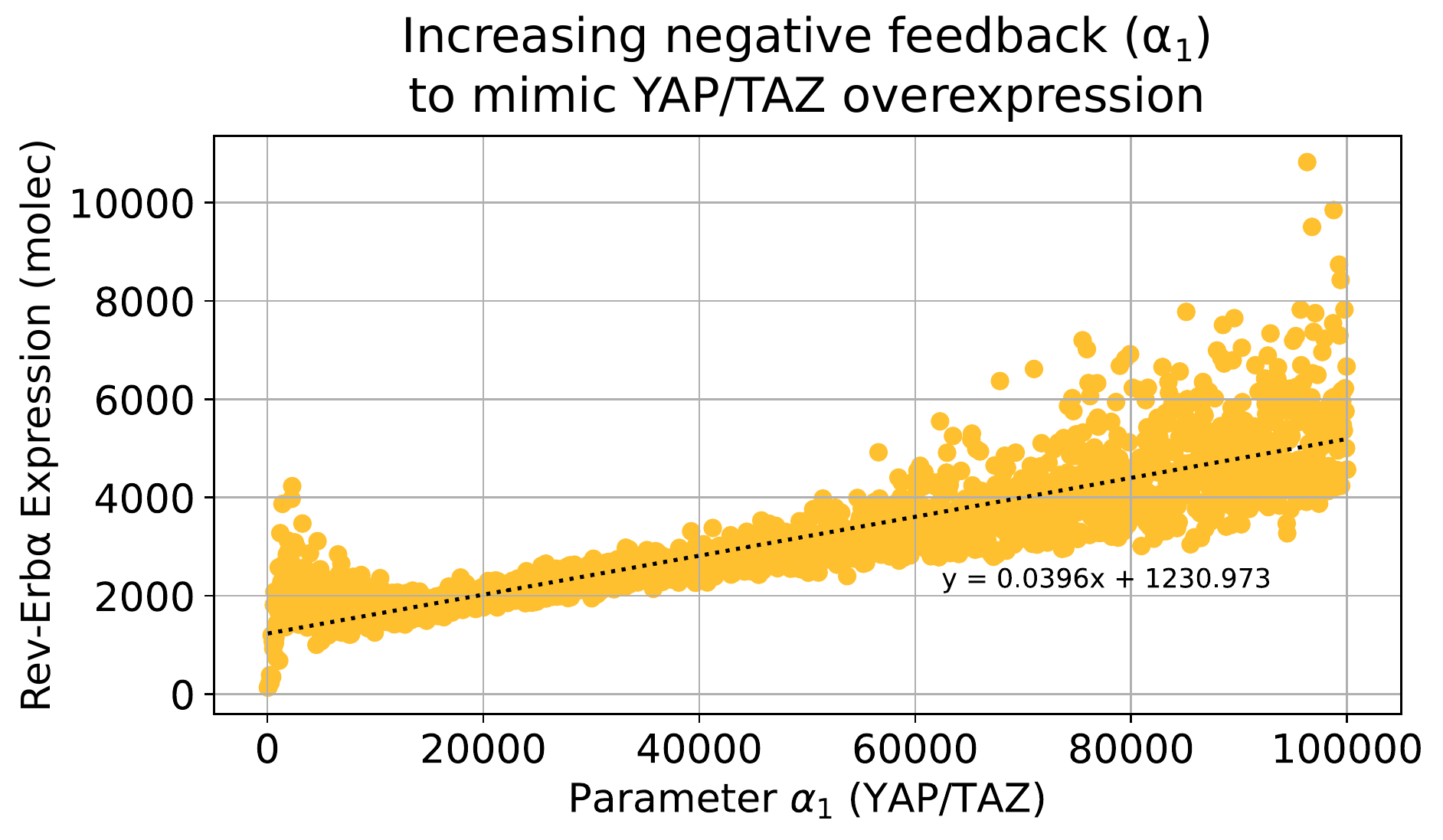}
	\caption{Effect of a continuous increase in $\alpha_1$ on Rev-Erb$\alpha$ levels.}
	\label{fig:revyap_scatter}
\end{figure}
\begin{figure*}[htbp]
	\centering
	\includegraphics[width=0.95\textwidth]{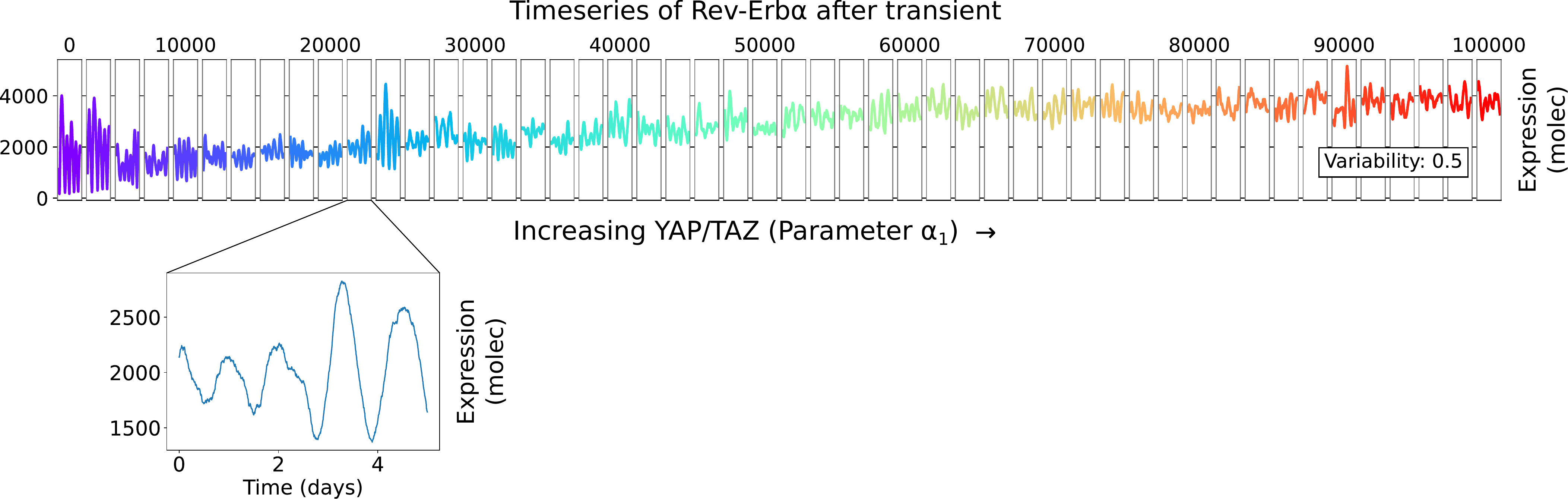}
	\caption{Stochastic trajectories as parameter $\alpha_1$ is increased.
	Each window shows a distinct time series with its own set of parameters.
	All model parameters, except for the Hill coefficient $h$, are chosen from a log-normal distribution with a coefficient of 0.5 (as measured from the standard deviation and mean of the underlying normal distribution in each case.
	The distributions are centered around the deterministic parameter values, and the standard deviation was adjusted to introduce greater variability.}
	\label{fig:rainbow_a}
\end{figure*}

\paragraph{Ensembles of stochastic simulations.}
Two groups of stochastic simulations are run, reflecting different concentrations of YAP/TAZ: low ($\alpha_1 = 5500$~molec$\cdot$h$^{-1}$) and high ($\alpha_1 = 10^6$~molec$\cdot$h$^{-1}$).
We recall that the level of YAP/TAZ is inversely related to cell density, where high cell density corresponds to low YAP/TAZ concentrations and vice versa.
For each case, 1000 simulations are run for 20 days, using the corresponding value of $\alpha_1$.
The remaining parameters are varied randomly with each simulation, adding an extra layer of noise to the system.
Specifically, the parameters are selected from a log-normal distribution, centered around the deterministic parameter values from Table \ref{tab:param_vals}, and with a spread that is roughly 20\% the parameter mean. 

\paragraph{Stochastic time series.}
The stochastic time series, examples of which are shown in Fig.~\ref{fig:sample_timeseries}, display steady oscillations for the case of low YAP/TAZ (high cell density), with a period around 24 hours on average. The amplitude varies greatly between cycles of a given time series, but the period remains fairly consistent. For high YAP/TAZ (low cell density), the period was higher than 24 hours for most of the simulations. Both the amplitude and period change from cycle to cycle, and the modulation depth is much lower compared to the high cell density case.
Kymographs of the time series are shown in Fig.~\ref{fig:kymograph_simulations}(a). Now that noise has been added to the system, the model better reflects the results seen in our reference experiments \cite{Abenza2022}.

\paragraph{Fourier analysis.}
Fourier spectra were calculated and averaged over the 1000 simulations described above, for each of the two $\alpha_1$ values (Fig.~\ref{fig:kymograph_simulations}b).
When $\alpha_1 = 5500$~molec$\cdot$h$^{-1}$ (low YAP/TAZ, high cell density), the stochastic simulations have a dominant frequency centered at roughly 0.9~day$^{-1}$ (26.7~h), which disappears when the parameter $\alpha_1$ is increased to $10^6$~molec$\cdot$h$^{-1}$. These results are consistent with the experimental observations shown in Fig.~\ref{fig:exp} above.

\paragraph{Effect on circadian gene levels.}
As discussed briefly in Sec.~\ref{sec:model} above, in our deterministic model the average concentration of Rev-Erb$\alpha$ increases with $\alpha_1$.
The stochastic simulations show that not only does the concentration increase, but it also becomes more variable across the simulations.
This is shown in Fig.~\ref{fig:kymograph_simulations}(c), which plots the instantaneous levels of Rev-Erb$\alpha$ obtained in our simulations for both low and high YAP/TAZ levels.
These results fit with our experimental observations, where higher concentrations of TAZ correspond to higher Rev-Erb$\alpha$ concentrations with greater spread (Fig.~\ref{fig:exp}c) \cite{Abenza2022}.

The response of Rev-Erb$\alpha$ to a continuous increase of $\alpha_1$ is shown in Fig.~\ref{fig:revyap_scatter}, ranging from $\alpha_1=500$~molec$\cdot$h$^{-1}$ to $10^5$~molec$\cdot$h$^{-1}$, with steps of $50$~molec$\cdot$h$^{-1}$.
For each value of $\alpha_1$, 10 simulations were run for 10 days. This allows us to see the behavior of the model throughout the entire range of YAP/TAZ levels.
As the parameter $\alpha_1$ increases sufficiently, the expression of Rev-Erb$\alpha$ increases overall.
For small values of $\alpha_1$ (less than $2\cdot 10^4$~molec$\cdot$h$^{-1}$), the average concentration increases first and then decreases, as $\alpha_1$ passes through and exits the oscillatory region.
After this, the concentration steadily increases.
Additionally, the average expression becomes more variable the further $\alpha_1$ is increased.
The results shown in Fig.~\ref{fig:revyap_scatter} do not take into account the time structure of the system.
For completeness, Fig.~\ref{fig:rainbow_a} shows how the time series behave as $\alpha_1$ increases continuously. The results show that circadian oscillations persist beyond the regime of deterministic limit cycle oscillations \cite{neiman97}.

\section{Discussion}

Here we used a modified version of the Goodwin model to simulate a circadian clock.
The effect of the mechanotransducer YAP/TAZ on the robustness of the clock was incorporated by varying the parameter associated with the strength of negative feedback in Bmal1.
Overall, the results from the model are consistent with experimental results \cite{Abenza2022}.
As the concentration of YAP/TAZ increased, the average concentration of Rev-Erb$\alpha$ increased as well.
Overall, the results of the model support the idea that YAP/TAZ concentrations affect the circadian behavior of mammalian cells via its connection to Bmal1.
The model also predicts that circadian oscillations would be lost for low values of $\alpha_1$, corresponding to cell densities higher than the basal values considered in the experiments shown in Fig.~\ref{fig:exp}.

Our study required us to modify the original Goodwin model. 
Although the standard version of this model (in which all decay terms are linear) indeed describes an oscillatory system, it fails to capture all of the aspects of the experiments.
The principal issue of the standard model is that the oscillations continue to be present for increasing values of $\alpha_1$.
Additionally, the period of the oscillations stays fixed at 24 hours no matter how high $\alpha_1$ is raised. 
Another issue with the standard model is the modulation depth of the time series, defined as the amplitude of the oscillations divided by their mean, which provides an idea of how substantial the oscillations are with respect to the average concentration values.
Here we have shown that these issues can be addressed by changing the linear degradation term for Rev-Erb$\alpha$ protein to include a saturation term (Eq.~\ref{eqn:goodwin}).
Notably, this also allows for a much smaller Hill coefficient $h$ in the negative feedback term of $X$ by $Z$ \cite{murray_book}, which can now be decreased from 10 to 2.

There are other ways in which the experimental results of Fig.~\ref{fig:exp} could be reproduced with our model.
Specifically, a dimensional analysis shows that the model could be rewritten in terms of a smaller number of free parameters as:
\begin{eqnarray}
	&\frac{dX'}{dt'}& = \frac{\alpha_1'}{1+(\frac{Z}{K'}) ^{h}} - X, \\
	&\frac{dY'}{dt'}& = X - d_2'Y, \\
	&\frac{dZ'}{dt'}& = Y - \frac{d_3'Z}{1+Z},
	\label{eqn:goodwin2}
\end{eqnarray}
where all variables and parameters are now dimensionless. In particular $\alpha_1'=\alpha_1\beta_2\beta_3/(Sd_1^3)$.
This shows that the increase in $\alpha_1$ studied above can be recapitulated by an increase in $\beta_2$ and/or in $\beta_3$.
Thus we cannot discard with our analysis that the effect of YAP/TAZ could take place via the transcription or translation of Rev-Erb$\alpha$ directly, which should be considered alternative predictions of our model.
In any case, we chose to interpret our results in terms of mechanical effects on Bmal1, on the basis of the above-mentioned experimental observations that YAP/TAZ influences that clock protein  \cite{Zhao2008,Zanconato2015,Lee2016,Rivera-Reyes2018,Rajbhandari2018}.

Further work could be done to consider how cells behave in proximity to other cells through the use of agent-based modeling to model mechanical effects among fibroblasts.
Each cell would contain its own circadian model, regulating the concentrations of Bmal1 and Rev-Erb$\alpha$.
Changes in the cellular environment (high or low cell density) would drive changes in the parameter $\alpha_1$. 
This could easily be adapted to respond to other mechanical inputs into the cell, potentially shedding further light on the interplay between the mechanical and biochemical regulation of cellular dynamics.

\section*{Acknowledgements}

This work was supported by project PID2021-127311NB-I00 financed by the Spanish Ministry of Science and Innovation, the Spanish State Research Agency and FEDER (MICIN/AEI/10.13039/501100011033/FEDER), by the Maria de Maeztu Programme for Units of Excellence in R\&D (project CEX2018-000792-M), and by the Generalitat de Catalunya (ICREA Academia programme). 

\bibliography{yap}

\end{document}